\documentstyle[12pt]{article}
\advance \topmargin by -\headheight
\advance \topmargin by -\headsep

\evensidemargin \oddsidemargin
\begin{document}

\topmargin -.6in

\def\rf#1{(\ref{eq:#1})}
\def\lab#1{\label{eq:#1}}
\def\nonu{\nonumber}
\def\br{\begin{eqnarray}}
\def\er{\end{eqnarray}}
\def\be{\begin{equation}}
\def\ee{\end{equation}}
\def\eq{\!\!\!\! &=& \!\!\!\! }
\def\lb{\lbrack}
\def\rb{\rbrack}
\def\llangle{\left\langle}
\def\rrangle{\right\rangle}
\def\blangle{\Bigl\langle}
\def\brangle{\Bigr\rangle}
\def\llbrack{\left\lbrack}
\def\rrbrack{\right\rbrack}
\def\lcurl{\left\{}
\def\rcurl{\right\}}
\def\({\left(}
\def\){\right)}
\newcommand{\nit}{\noindent}
\newcommand{\ct}[1]{\cite{#1}}
\newcommand{\bi}[1]{\bibitem{#1}}
\def\lskip{\vskip\baselineskip\vskip-\parskip\noindent}
\relax

\def\v{\vert}
\def\bv{\bigm\vert}
\def\Bgv{\;\Bigg\vert}
\def\bgv{\bigg\vert}
\newcommand\partder[2]{{{\partial {#1}}\over{\partial {#2}}}}
\newcommand\funcder[2]{{{\delta {#1}}\over{\delta {#2}}}}
\newcommand\Bil[2]{\Bigl\langle {#1} \Bigg\vert {#2} \Bigr\rangle}  
\newcommand\bil[2]{\left\langle {#1} \bigg\vert {#2} \right\rangle} 
\newcommand\me[2]{\left\langle {#1}\bv {#2} \right\rangle} 
\newcommand\sbr[2]{\left\lbrack\,{#1}\, ,\,{#2}\,\right\rbrack}
\newcommand\pbr[2]{\{\,{#1}\, ,\,{#2}\,\}}
\newcommand\pbbr[2]{\lcurl\,{#1}\, ,\,{#2}\,\rcurl}
%
\def\a{\alpha}
\def\b{\beta}
\def\dc{{\cal D}}
\def\d{\delta}
\def\D{\Delta}
\def\eps{\epsilon}
\def\vareps{\varepsilon}
\def\g{\gamma}
\def\G{\Gamma}
\def\grad{\nabla}
\def\h{{1\over 2}}
\def\l{\lambda}
\def\L{\Lambda}
\def\m{\mu}
\def\n{\nu}
\def\o{\over}
\def\om{\omega}
\def\O{\Omega}
\def\p{\phi}
\def\P{\Phi}
\def\pa{\partial}
\def\pr{\prime}
\def\ra{\rightarrow}
\def\s{\sigma}
\def\S{\Sigma}
\def\t{\tau}
\def\th{\theta}
\def\Th{\Theta}
\def\ti{\tilde}
\def\wti{\widetilde}
\def\bj{{\bar J}}
\def\sj{{\jmath}}
\def\bsj{{\bar \jmath}}
\def\bp{{\bar \p}}
\newcommand\sumi[1]{\sum_{#1}^{\infty}}   
\newcommand\fourmat[4]{\left(\begin{array}{cc}  
{#1} & {#2} \\ {#3} & {#4} \end{array} \right)}
\def\gi{g^{-1}\,}
\def\gh{g^{\h}\,}
\def\ghi{g^{-\h}\,}
%
\def\lie{{\cal G}}
\def\dlie{{\cal G}^{\ast}}
\def\elie{{\widetilde \lie}}
\def\edlie{{\elie}^{\ast}}
\def\hlie{{\cal H}}
\def\wlie{{\widetilde \lie}}
\def\winf{{\sf w_\infty}}
\def\win1{{\sf w_{1+\infty}}}
\def\hwinf{{\sf {\hat w}_{\infty}}}
\def\Winf{{\sf W_\infty}}
\def\Win1{{\sf W_{1+\infty}}}
\def\hWinf{{\sf {\hat W}_{\infty}}}
%
\def\rlx{\relax\leavevmode}
\def\inbar{\vrule height1.5ex width.4pt depth0pt}
\def\IZ{\rlx\hbox{\sf Z\kern-.4em Z}}
\def\IR{\rlx\hbox{\rm I\kern-.18em R}}
\def\IC{\rlx\hbox{\,$\inbar\kern-.3em{\rm C}$}}
%
\def\mark{\noindent{\bf Remark.}\quad}
\def\prop{\noindent{\bf Proposition}\quad}
\def\proof{\noindent{\bf Proof.}\quad}
\def\um{linear two-boson KP hierarchy }
\def\dois{quadratic two-boson KP hierarchy }
\def\treis{cubic two-boson KP hierarchy }
\def\uma{linear two-boson KP hierarchy}
\def\doisa{quadratic two-boson KP hierarchy}
\def\treisa{cubic two-boson KP hierarchy}
%
%
\def\PRL#1#2#3{{\sl Phys. Rev. Lett.} {\bf#1} (#2) #3}
\def\NPB#1#2#3{{\sl Nucl. Phys.} {\bf B#1} (#2) #3}
\def\NPBFS#1#2#3#4{{\sl Nucl. Phys.} {\bf B#2} [FS#1] (#3) #4}
\def\CMP#1#2#3{{\sl Commun. Math. Phys.} {\bf #1} (#2) #3}
\def\PRD#1#2#3{{\sl Phys. Rev.} {\bf D#1} (#2) #3}
\def\PLA#1#2#3{{\sl Phys. Lett.} {\bf #1A} (#2) #3}
\def\PLB#1#2#3{{\sl Phys. Lett.} {\bf #1B} (#2) #3}
\def\JMP#1#2#3{{\sl J. Math. Phys.} {\bf #1} (#2) #3}
\def\PTP#1#2#3{{\sl Prog. Theor. Phys.} {\bf #1} (#2) #3}
\def\SPTP#1#2#3{{\sl Suppl. Prog. Theor. Phys.} {\bf #1} (#2) #3}
\def\AoP#1#2#3{{\sl Ann. of Phys.} {\bf #1} (#2) #3}
\def\PNAS#1#2#3{{\sl Proc. Natl. Acad. Sci. USA} {\bf #1} (#2) #3}
\def\RMP#1#2#3{{\sl Rev. Mod. Phys.} {\bf #1} (#2) #3}
\def\PR#1#2#3{{\sl Phys. Reports} {\bf #1} (#2) #3}
\def\AoM#1#2#3{{\sl Ann. of Math.} {\bf #1} (#2) #3}
\def\LMP#1#2#3{{\sl Letters in Math. Phys.} {\bf #1} (#2) #3}
\def\IJMPA#1#2#3{{\sl Int. J. Mod. Phys.} {\bf A#1} (#2) #3}
\def\TMP#1#2#3{{\sl Theor. Mat. Phys.} {\bf #1} (#2) #3}
\def\JPA#1#2#3{{\sl J. Physics} {\bf A#1} (#2) #3}
\def\MPLA#1#2#3{{\sl Mod. Phys. Lett.} {\bf A#1} (#2) #3}
\def\JETP#1#2#3{{\sl Sov. Phys. JETP} {\bf #1} (#2) #3}
\def\JETPL#1#2#3{{\sl  Sov. Phys. JETP Lett.} {\bf #1} (#2) #3}
\def\PHSA#1#2#3{{\sl Physica} {\bf A#1} (#2) #3}
\def\PHSD#1#2#3{{\sl Physica} {\bf D#1} (#2) #3}

\begin{titlepage}
\vspace*{-1cm}
\noindent
July, 1993 \hfill{IFT-P.041/93}\\
\phantom{bla}
\hfill{hep-th/9307147}
\\
\vskip .3in

\begin{center}

{\large\bf Toda and Volterra Lattice Equations from }
\end{center}
\begin{center}
{\large\bf Discrete Symmetries of KP Hierarchies}
\end{center}
\normalsize
\vskip .4in

\begin{center}
{ H. Aratyn\footnotemark
\footnotetext{Work supported in part by U.S. Department of Energy,
contract DE-FG02-84ER40173 and by NSF, grant no. INT-9015799}}

\par \vskip .1in \noindent
Department of Physics \\
University of Illinois at Chicago\\
845 W. Taylor St.\\
Chicago, Illinois 60607-7059\\
\par \vskip .3in

\end{center}

\begin{center}
{L.A. Ferreira\footnotemark
\footnotetext{Work supported in part by CNPq}}, J.F. Gomes$^{\,2}$
and A.H. Zimerman$^{\,2}$

\par \vskip .1in \noindent
Instituto de F\'{\i}sica Te\'{o}rica-UNESP\\
Rua Pamplona 145\\
01405-900 S\~{a}o Paulo, Brazil
\par \vskip .3in

\end{center}

\begin{center}
{\large {\bf ABSTRACT}}\\
\end{center}
\par \vskip .3in \noindent

The discrete models of the Toda and Volterra chains are being constructed
out of the continuum two-boson KP hierarchies.
The main tool is the discrete symmetry preserving the Hamiltonian
structure of the continuum models.
The two-boson currents of KP hierarchy are being associated with
sites of the corresponding chain by successive actions of discrete
symmetry.
\end{titlepage}
{\large {\bf 1. Introduction}}
\lskip
One of the challenging problems of the multi-matrix models
is extraction of the continuum differential hierarchies from
lattice hierarchies one encounters in these models.
In a series of papers \ct{BX9204,BX9209,BX9212} a new proposal, different
from the usual double scaling approach, was put up.
It builds on a observation, which takes a specially simple form in the
context of one-matrix model. There the relevant system is given
by a spectral equation of the Toda chain.
This equation can be rewritten in terms of pseudo-differential
operator of the type one encounters in the study of continuum two-boson
KP systems \ct{BX9204,2boson}. Technically the trick one uses in this
process involves taking equations of motion in order to rewrite the
spectral equation in terms of fields defined on one lattice site only.
One can then view this result as an indication that the lattice system can be
decomposed in many disjoint but isomorphic KP hierarchies each associated
with the separate lattice sites.

In this paper we uncover the connection between on one side the discrete
Toda, Volterra and modified Volterra models \ct{W76,KM92} and on another
three continuum, integrable systems belonging to the class of two-boson
KP hierarchies \ct{YW9111,BAK85,2boson}.
As noticed before \ct{AFGMZ} in case of one of these continuum models
there exists in the two-boson KP models a new type of fundamental,
discrete symmetry.
Here we describe its role and properties in the larger context of three
basic two-boson KP hierarchies.
Furthermore we use this discrete symmetry to build the corresponding lattice
models where it is realized as translation between neighboring sites.
Due to canonical nature of this symmetry we can associate to each site
the isomorphic KP system.
Hence this part of our findings confirms early predictions of ideas
contained in \ct{BX9204}.
On the other hand the symmetry itself can be used to directly build
the site objects out of the continuum objects in such a way that the
continuum models appear to be solutions of the lattice equations.

The picture, which emerges from this analysis of the integrable systems
considered by us, bears some resemblance to the concept of (trivial)
fiber bundle framework in gauge theories.
It is natural to associate the discrete symmetries and therefore translations
between neighboring lattice sites with vertical direction along the fiber.
For each site we can project on the equivalent copy of the continuum
two-boson KP model.
In this analogy the dynamics of the continuum KP model and that of its
discrete counterpart take place in the orthogonal directions.
Kinematics of the lattice system gives rise to symmetry operation on
the continuum model.
In this sense the continuum model due to its intrinsic discrete symmetry
can be viewed as a solution of the lattice equations.

While working on this paper we have learned that authors of \ct{LSY93}
are also aware of the connection between translations on the Toda lattice and
the discrete symmetries of continuum integrable systems.
\lskip
{\large {\bf 2. Symmetries of Two-Boson KP Systems}}
\lskip
Here we present the main two-boson KP systems with an emphasis on properties
allowing for establishing correspondence with lattice systems.
It turns out that from the point of view of constructing the discrete versions
of the two-boson KP models the main role is being played by
the canonical transformations preserving Poisson and Hamiltonian
structures.
There are two types of such canonical mappings.
The generalized Miura transformations map various two-boson KP hierarchies
into each other \ct{AFGMZ}.
They are in general associated with gauge transformations acting
simply between Lax operators and possessing property of being
canonical (symplectic) with Hamiltonians of one model being transformed
into Hamiltonians of another one \ct{ANPV}.
The Miura transformations will remain valid on the lattice connecting
various discrete models.

Then there are discrete canonical mappings which act inside each hierarchy
\ct{AFGMZ}.
They are associated with translations on the corresponding lattice systems
and are invertible. They preserve the Poisson and Hamiltonian structure
of the two-boson KP hierarchies.

Because of simplicity of the discrete canonical mapping in the setting
of the \dois we start with a short review of this model.

\nit {\sl Quadratic Two-Boson KP Hierarchy.}
Here the pseudo-differential operator is \ct{YW9111}:
\be
L_{\sj} = D + \bsj\, \(D - \sj \,- \bsj \,\)^{-1} \sj
\lab{sjlax}
\ee
Among the corresponding Poisson structures only second and third are local
and are given by \ct{depir}:
\be
P_{2} \lb \sj\; \rb = \left(\begin{array}{cc}
0 & D \\
D & \; 0 \end{array}
\right) \;\; , \;\;
P_{3} \lb \sj\; \rb  =\left(\begin{array}{cc}
D \sj \,+ \sj\, D & \; -D^2 + D \sj\, + \bsj \, D \\
D^2 + \sj\, D+ D \bsj \, &\; D\bsj\, + \bsj \, D \end{array}
\right)
\lab{p1sjp2sj}
\ee
In terms of $P_{i} \lb \sj\; \rb $ we can express the $i$-th bracket as
\be
\{ A\, , \, B \}_i = \( {\d A \o \d \sj\,} \; {\d A \o \d \bsj\,} \)\;
P_{i} \lb \sj\; \rb \;
{\d B / \d \sj \choose  \d B/ \d \bsj\, }
\lab{abbra}
\ee
The flow equations read in terms of Hamiltonians $H_r \lb \sj\, \rb$ as:
\be
{ {\d \sj\,}/ {\d t_r} \choose {\d \bsj\,}/ {\d t_r} }
=  P_{1} \lb \sj\; \rb { {\d H_{r+1} \lb \sj\, \rb }/ {\d \sj\,} \choose
{\d H_{r+1}\lb \sj\, \rb }/ {\d \bsj\,}} =
P_{2} \lb \sj\; \rb { {\d H_{r} \lb \sj\, \rb}/ {\d \sj\,} \choose
{\d H_{r}\lb \sj\, \rb }/ {\d \bsj\,}}
= P_{i} \lb \sj\; \rb \, { {\d H_{r+2-i} \lb \sj\, \rb}/ {\d \sj\,} \choose
{\d H_{r+2-i}\lb \sj\, \rb }/ {\d \bsj\,}}
\lab{sjflow}
\ee
For the lowest flow we find easily
\be
{\pa \sj\,\o \pa t_1} = {\pa \sj\,\o \pa x} \equiv \sj^{\,\pr} \qquad; \qquad
{\pa \bsj\,\o \pa t_1} = {\pa \bsj\,\o \pa x} \equiv \bsj^{\,\pr}
\lab{sjflow1}
\ee
Compatibility relation between various Hamiltonian structures
can be put in the form of recurrence relation:
\be
P_i \lb \sj\; \rb = P_1 \lb \sj\; \rb \((P_1 \lb \sj\; \rb )^{-1}
P_2 \lb \sj\; \rb \)^{i-1} = P_{i-1} \lb \sj\; \rb (P_1 \lb \sj\; \rb )^{-1}
P_2 \lb \sj\; \rb
\qquad\; i\geq 1 \lab{pip1p2}
\ee
Hence we can express any multi-Hamiltonian structure in the
quadratic two-boson hierarchy
in terms of only $P_{2}\lb \sj\; \rb $ and $P_{3}\lb \sj\; \rb $ by
recalling that $P_{3} \lb \sj\; \rb = P_{2}\lb \sj\; \rb
P_{1}^{-1} \lb \sj\; \rb P_{2}\lb \sj\; \rb $.

The three lowest Hamiltonian functions are
$H_{1} \lb \sj\; \rb \! = \! \! \int \! \sj\, \bsj \;$,
$H_{2} \lb \sj\; \rb \! = \! \! \int \! - \sj^{\pr}\, \bsj\, + \sj^{\,2}\,
\bsj +\sj\, \bsj^{\,2}$ and $H_3 \lb \sj\,\rb \! = \! \! \int \!
\sj^{\pr \pr}\, \bsj - 3 \sj \,\sj^{\pr}\, \bsj
- 2 \sj^{\pr}\, \bsj^2 - \sj\, \bsj\, \bsj^{\,\pr} + \sj^{3} \, \bsj
+ 3 \sj^{2} \, \bsj^{\,2} + \sj\, \bsj^{\,3}$.

One of the main points of this section is the existence of an infinite,
discrete transformation group $Z_g$ (isomorphic to integers), which keeps
the Hamiltonian structure of the \dois invariant.
This transformation group is generated by substitutions \ct{AFGMZ}:
\br
g (\sj\,)  & \equiv & \bsj\, - {\sj^{\,\pr} \o \sj\,} \qquad \quad \quad
\mbox{\rm and}
\qquad \quad \quad g\;( \bsj\,) \equiv \sj
\lab{sjtransf} \\
\gi (\bsj\,) &\equiv& \sj\, + { \bsj^{\,\pr} \o \bsj \, }\qquad \quad\quad
\mbox{\rm and}    \qquad \quad \quad \gi ( \sj\,) \equiv \bsj
\lab{sjtransf2}
\er
\prop $\!\!${\bf 1.}~ The Hamiltonian structure corresponding to the Lax
operator $L_{\sj}$ in \rf{sjlax} is invariant under $Z_g$.

\proof One verifies easily that the bracket structures induced by
 both $P_{2} \lb \sj\; \rb $ and $P_{3}\lb \sj\; \rb $ are invariant under
the substitutions \rf{sjtransf} and \rf{sjtransf2}, meaning that
\be
\fourmat{{\pbr{g(\sj\,)}{g(\sj\,)}}_i}{{\pbr{g(\sj\,)}{g(\bsj\,)}}_i}
{{\pbr{g(\bsj\,)}{g(\sj\,)}}_i}{{\pbr{g(\bsj\,)}{g(\bsj\,)}}_i}
=  P_{i} \lb g (\sj\;) \rb \qquad  {\rm for \;} \quad i=2,3
\lab{proofa}
\ee
with $g$ acting on both $\sj\,$ and $\bsj\,$.
Recalling that $P_1 \lb \sj\; \rb = P_2 \lb \sj\; \rb P_3^{-1} \lb \sj\;\rb
P_2 \lb \sj\; \rb $, it is easy to convince oneself that
a recurrence matrix $P_2 \lb \sj\; \rb (P_1 \lb \sj\; \rb )^{-1}$ as well as
all remaining higher Poisson structures must therefore remain invariant
under substitutions \rf{sjtransf} and \rf{sjtransf2}.
Hence $g$ is an automorphism of all Poisson structures satisfying a relation:
\be
g \( {\pbr{A}{B}}_r \) = {\pbr{g(A)}{g(B)}}_r  \qquad \qquad r= 1,2,\ldots
\lab{auto}
\ee
with $A$ and $B$ being some arbitrary polynomials in $\sj\,$ and $\bsj\,$.
The rest of the proof goes by induction based on the Lenard relation
$ {\pbr{A}{H_{r+1}}}_2 = {\pbr{A}{H_{r}}}_3$.
We first make an assumption that $ g \( H_{r} \lb \sj\; \rb \)
= H_{r} \lb \sj\; \rb$.
{}From \rf{auto} we have:
\be
{\pbr{g(A)}{g(H_{r+1})}}_2 = {\pbr{g(A)}{g(H_r)}}_3 =
{\pbr{g(A)}{H_r}}_3 = {\pbr{g(A)}{H_{r+1}}}_2
\lab{induction}
\ee
Hence it follows that $ {\pbr{g(A)}{g(H_{r+1})-H_{r+1}}}_2$ for an
arbitrary $A$.
Especially, recalling the form of the bracket ${\pbr{\cdot}{\cdot}}_2$
from \rf{p1sjp2sj} we have
\be
\pa_x {\d \lb g(H_{r+1})-H_{r+1} \rb\o \d \sj\,} = 0 \qquad; \qquad
\pa_x { \d \lb g(H_{r+1})-H_{r+1} \rb \o \d \bsj\,} = 0
\lab{indend}
\ee
Hence if $H_{r} \lb \sj\; \rb$ is invariant, $H_{r+1} \lb \sj\; \rb $ is
invariant too.
Since one verifies easily by inspection that
$H_{1} \lb \sj\; \rb = g \( H_{1} \lb \sj\; \rb\) $ this completes the proof.

\nit {\sl Linear two-boson KP Hierarchy.}
The \um is defined in terms of the Lax operator:
\be
L_{J} = D -J + \bj\, D^{-1}
\lab{Jlax}
\ee
The corresponding multi-Hamiltonian structures have been studied in
\ct{BAK85,2boson}. This system contains three local Poisson structures
$P_{i} \lb J \rb \; i=1,2,3$.
We can express any Poisson structure $P_{i} \lb J \rb \;,\; i=3,4,\ldots$
by $P_{2} \lb J \rb $ and $P_{1} \lb J \rb $ through recurrence relation
$P_{i} \lb J \rb =\(P_{2} \lb J \rb (P_{1})^{-1} \lb J \rb \)^{i-2}
P_{2} \lb J \rb $ involving the recurrence matrix
$P_{2} \lb J \rb (P_{1})^{-1} \lb J \rb $ \ct{2boson}.

The explicit form of first and second local Hamiltonian structures is:
\be
P_{1} \lb J \rb = \left(\begin{array}{cc}
0 & - D \\
-D & \; 0 \end{array}
\right) \;\; , \;\;
P_{2} \lb J \rb =\left(\begin{array}{cc}
2 D & \; D^2 + D J \\
- D^2 + J D &\; D \bj+ \bj D\end{array}
\right)
\lab{Jp1Jp2}
\ee
The two lowest Hamiltonian functions are $H_1 \lb  J \rb
= \int \bj $ and $H_{2} \lb  J \rb = - \int \bj J$
with the lowest flow equation taking again the form ${\pa J}/{\pa t_1}
= J^{\pr} \; ; \; {\pa \bj}/{\pa t_1} = \bj^{\pr}$.

As in the \dois there exists the discrete transformation group $Z_G$
with generators defined as:
\br
G (J)  & \equiv & J + \( \ln \( \bj + J^{\pr} \) \)^{\pr} \qquad
\quad \mbox{\rm and}
\qquad \quad \quad \; G ( \bj) \equiv \bj + J^{\pr}
\lab{jtransf} \\
G^{-1} (J) &\equiv& J - \( \ln \bj \)^{\pr}  \qquad \quad\quad\quad\quad
\mbox{\rm and}    \qquad \quad \quad G^{-1} (\bj) \equiv
\bj + \( \ln \bj  \)^{\pr \pr} - J^{\pr}
\lab{jtransf2}
\er
Again action of $Z_G$ leaves the corresponding
Hamiltonian structure of the \um invariant .
This will follow from the Proposition 1 once
we establish below a relation between the linear and quadratic two-boson
KP hierarchies.

\nit {\sl Gauge Equivalence between Linear and Quadratic Two-Boson
Hierarchies. Generalized Miura Map.}
Here we will show that the generalized Miura map connecting above two
hierarchies takes the form of gauge transformation \ct{AFGMZ}, which
preserves the bracket structure \ct{ANPV}.
Let us transform the Lax operator $L_{\sj}$ from \rf{sjlax} by the
gauge transformation generated by $\xi  = \( \int (\sj\, +\bsj\,)
- \ln \sj\,\)$:
\be
L_{\sj} \to \exp ( - \xi )\,L_{\sj}\, \exp (\xi ) = D + \sj \,+\bsj \, +
\sj \, ( \sj^{-1} \, )^{\pr} \, +  \bsj \, \sj \, D^{-1}
= D - J + \bj D^{-1}
\lab{gaugea}
\ee
where we have introduced
\be
J = - \sj \,- \bsj \, + {\sj^{\,\pr}  \o \sj}  \qquad ; \qquad
\bj = \bsj \, \sj            \lab{miura1}
\ee
One can now verify explicitly that with the bracket structure given by
$P_{2} \lb \sj\, \rb$ in \rf{p1sjp2sj} variables defined in \rf{miura1}
satisfy the second bracket structure $P_{2}\lb J \rb$ \rf{Jp1Jp2} of linear
two-boson KP hierarchy.
Relation \rf{miura1} represents a Miura transform for two-Bose hierarchies,
which generalizes the usual Miura transformation
between one-Bose KdV and mKdV structures \ct{AFGMZ}.
Note, that the existence of the transformation group $Z_g$ introduces an
ambiguity in a possible form of generalized Miura transformation and
\rf{miura1} can also take other equivalent forms.
Consider for instance:
\be
J = -\sj \,-\,\bsj \, \qquad ; \qquad
\bj = \bsj\, \sj \,+ \bsj^{\,\pr} \lab{miura2}
\ee
It is easy to see that the substitution $g$ takes \rf{miura2} into \rf{miura1}
and ensures therefore the canonical character of \rf{miura2}.

Miura transformation can be used to explain both the form of the
discrete mapping $G$ and its canonical property.
It follows namely from \rf{sjtransf} and \rf{miura1}
that
\be
g^{2} (J)  = g^2 \( - \sj \,- \bsj \, + {\sj^{\,\pr}  \o \sj\,}  \)
=  J + \( \ln \( \bj + J^{\pr} \) \)^{\pr}  = G (J) \lab{gsquare}
\ee
where we used once more the identification between two systems provided
by \rf{miura1}. The same identity holds for $\bj$ showing that
$G=g^2 $ when quantities of \um and \dois are connected via generalized Miura
transformations.

\nit {\sl The Cubic Two-boson KP Hierarchy.}
We first establish link between the \dois and the KP hierarchy associated
with so-called derivative Non-Linear Schr\"odinger (dNLS) system.
This connection originates from the gauge transformation:
\br
e^{\int (-\sj\, +\bsj\,) } \(D + \sj \,+ \bsj\, + \bsj \, D^{-1} \sj \)
e^{\int (\sj\, -\bsj\,) }\eq D + 2 \sj\, + \bsj\,
e^{\int (- \sj\, +\bsj\,) } D^{-1} \sj\, e^{\int (\sj\, -\bsj\,) } \nonu\\
\eq D + 2 r q + \( r q^2 + q^{\pr}\) D^{-1} r
\lab{sjtodnls}
\er
with the Miura map between variables of the \dois and dNLS hierarchy
taking form
\be
\sj \,(x) = q(x)r(x) \quad\; ; \quad \;
\bsj\, (x) = q(x)r(x) +{q^{\pr}(x) \o {q(x)}}= \sj\,(x) +
{q^{\pr}(x) \o {q(x)}}       \lab{j}
\ee
This Miura map is invertible and we find $q= \exp \int ( \bsj\,-\sj\,)$
and $r =\sj\, \exp - \int ( \bsj\,-\sj\,)$. {}From this relations we find
that $g (r) = q - r^{\pr} / r^2$ and $g (q) = r$.
We now propose the following third Poisson bracket structure
\be
\{q(x),r(y)\}_3 = \d^{\pr}(x-y) \quad \; ; \quad \;
\{q(x),q(y)\}_3 =  \{r(x),r(y)\}_3 = 0
\lab{br3}
\ee
We find that \rf{j} maps the structure given in eq. \rf{p1sjp2sj}
exactly to the above third bracket.
The corresponding equations of motion
\br
\dot q(x) = \{q(x), H_1\}_3 &=& q^{\pr \pr}(x) + 2(q^2(x)r(x))^{\pr} \nonu\\
\dot r(x) = \{r(x), H_1\}_3 &=& -r^{\pr \pr}(x) + 2(r^2(x)q(x))^{\pr}
\lab{eqmotion}
\er
correspond to the derivative NLS equations described in \ct{KauNew}.

The second  bracket of dNLS system is realized by:
\br
\{q(x),q(y)\}_2 &=& q(x)q(y)\epsilon (x-y) \quad;\quad
\{r(x),r(y)\}_2 = r(x)r(y)\epsilon (x-y)\nonu\\
\{q(x),r(y)\}_2 &=& \d (x-y) - r(y)q(x)\epsilon (x-y)
\lab{br2}
\er
and agrees with the second bracket of the \dois via the Miura map.

The dNLS hierarchy has a simpler structure than \treis we are about to
introduce.
The Miura map \rf{j} is invertible opposite to the Miura maps we have seen
in \rf{miura1} or \rf{miura2}. This will not be the case with the Miura map
between the \dois and \treisa.
In case of dNLS hierarchy one finds that due to increasing non-linearity of
corresponding Lax operator the only local bracket appears to be connected
with the third bracket structure.
All remaining brackets can be shown to be non-local.
This feature will also hold for the same reasons in the \treisa.
We introduce the new hierarchy by defining the Miura transformation between
the \dois and \treis as
\br
\sj\;(x) &=& \(1- q(x)\) \(1- r(x)\) \nonu\\
\bsj \; (x) &=& \(1+ q(x)\) \(1+ r(x)\)  + \( \ln (q (x) - 1) \)^{\pr}
\lab{jtorq}
\er
We can then define the \treis by a set of Hamiltonians $H_n (r,q)$
obtained from $H_{n} \lb \sj\,\rb $ by substitution \rf{jtorq}.
The third bracket structure of the \treis is still given by \rf{br3}
and maps under Miura \rf{jtorq} to the combination $\{ \cdot , \cdot \}_3
+ 4 \{ \cdot , \cdot \}_2$ of the \doisa.
Because of compatibility of these brackets the third Poisson bracket
structure of the \treis is well-defined and furthermore all
Hamiltonians $H_n (r,q)$ are in involution.
The discrete canonical mapping for the \treis will be denoted by
$\gh$ and acts as
\br
\gh (q)  & \equiv & r  \qquad \quad \qquad\quad\quad \quad
\mbox{\rm and}
\qquad \quad \quad \gh ( r) \equiv q - {r^{\pr} \o r^2 - 1}
\lab{ghtransf} \\
\ghi (q) &\equiv&  r + {q^{\pr} \o q^2 - 1}
\qquad \quad\quad
\mbox{\rm and}    \qquad \quad \quad \ghi (r) \equiv q
\lab{ghtransf2}
\er
One can show that $(\gh)^2 = g$ when acting on $\bsj\,$ and $\sj\,$ defined as
in \rf{jtorq}.
It is natural therefore to expect $\gh$ to be a canonical mapping with
respect to \rf{br3}.
We have explicitly verified that $\gh$ keeps invariant the three lowest
Hamiltonians obtained from \dois by substituting \rf{jtorq}.
\lskip
{\large {\bf 3. Toda and Volterra Chain Systems.}}
\lskip
{\sl Toda and Volterra chains.}
In this section we show that the dynamics given by the flow equation
\rf{sjflow1} of the \dois leads via discrete symmetries of $Z_g$ to
Volterra chain equations.
Similar observations hold for the remaining two two-boson KP hierarchies
leading to the Toda and the modified Volterra equations.
This construction yields on each lattice site a gauge copy of the
corresponding continuum two-boson KP system.

\nit \prop $\!\!\!${\bf 2.}~
The following equations hold in the \dois:
\br
\pa_t \,g^{2n} (\bsj\,) \eq \llbrack g^{2n-2} (\sj\,) - g^{2n} (\sj\,)
\rrbrack g^{2n} (\bsj\,) \lab{volta}\\
\pa_t\, g^{2n} (\sj\,) \eq \llbrack g^{2n} (\bsj\,) - g^{2n+2} (\bsj\,)
\rrbrack g^{2n} (\sj\,) \lab{voltb}
\er
where $t=t_1$.

\proof
It is a matter of simple algebra to verify the following identity
using \rf{sjtransf} and the flow equation \rf{sjflow1}:
\br
g^{2n+2} (\bsj\,)  \eq g^{2n} \(g^2(\bsj\,)\) =
g^{2n} \(\bsj\,- {\sj^{\,\pr} \o \sj\,} \) =
g^{2n} \(\bsj\,\) - {g^{2n} (\sj\,)^{\pr} \o g^{2n}( \sj\,)} \nonu \\
\eq  g^{2n} \(\bsj\,\) - {\pa_t g^{2n} (\sj\,) \o g^{2n}( \sj\,)}
\lab{proofb}
\er
This proves \rf{voltb} and essentially identical consideration leads to
the proof of \rf{volta}.

Introducing definitions
\be
A_n \equiv g^{2n} \(\bsj\,\) \qquad;\qquad
B_n \equiv g^{2n} \(\sj\,\) \qquad \qquad n \in \IZ
\lab{abn}
\ee
we can rewrite \rf{volta} and \rf{voltb} as:
\br
\pa_t\, A_n &=& A_n \( B_{n -1} - B_n \) \lab{aneq}\\
\pa_t\, B_n &=& B_n \( A_{n} - A_{n+1} \) \lab{bneq}
\er
Eqs. \rf{aneq} and \rf{bneq} can be casted in the more compact form of
Volterra equation \ct{W76,KM92}:
\be
\pa_t\, V_n \,=\,-  V_n \( V_{n +1} - V_{n-1} \) \lab{volterka}
\ee
by associating $A_n$ and $B_n$ to $V_n$ in the following way
\be
A_n= V_{2n-1} \qquad;\qquad B_n = V_{2n} \lab{anbnvn}
\ee
We have therefore constructed Volterra chain system out of discrete symmetry
of the \doisa.
We have seen in the previous section that the \dois was connected in terms of
generalized Miura transformation with remaining two two-boson KP hierarchies.
We will now address the question what lattice systems can be obtained
by the lattice version of the generalized Miura transformations
seen above.
Following relation \rf{abn} and definition \rf{miura1} it is natural
to define
\br
S_n &\equiv& g^{2n} \( -\sj\, -\bsj\, + {\sj^{\,\pr} \o \sj\,}\)
= -B_n - A_n + {\pa_t B_n \o B_n}= -B_n - A_{n+1} =
g^{2n+1} \( -\sj\, -\bsj\, \)                          \lab{lmiura1a}\\
R_n &\equiv& g^{2n} \( \bsj\, \sj\,\) = A_n B_n        \lab{lmiura1b}
\er
where use was made of \rf{bneq}.
Existence of the discrete symmetry $g$ introduced an ambiguity in
the way we could connect different two-boson KP hierarchies.
Example of this ambiguity is given by an alternative definition
of the Miura map in \rf{miura2}. Following this alternative
definition we can define
\br
{\wti S}_{n-1} &\equiv& g^{2n} \( -\sj\, -\bsj\, \)
= -B_n - A_n                         \lab{lmiura2a}\\
{\wti R}_{n-1} &\equiv& g^{2n} \( \bsj\, \sj\, + \bsj^{\,\pr} \) =
A_n B_n + \pa_t A_n =  A_n B_{n -1} = g^{2n-1} \( \bsj\, \sj\, \)
\lab{lmiura2b}
\er
It is easy to check that both set of variables satisfy
the equations of motion of the Toda chain:
\br
\pa_t \,S_n &= & R_{n +1} - R_n  \lab{sneq}\\
\pa_t\, R_n &=& R_n \( S_{n} - S_{n-1} \) \lab{rneq}
\er
Recalling definition of the discrete mapping $G$ in the \um
we can rewrite definitions \rf{lmiura1a} and \rf{lmiura1b}
as
\be
S_n = G^n \( J\) \qquad ;\qquad R_n = G^n \( \bj \) \lab{snrnj}
\ee
For completeness we mention here a definition of the Toda chain in terms of
$\tau$-functions. This involves the quantities
\be
\p_n \equiv \int G^n \( J\) dt \qquad;\qquad
\pa_t \ln \tau_n \equiv  \int G^n \( \bj \)
\lab{taudef}
\ee
which on basis of properties \rf{jtransf} and \rf{jtransf2} of $G$ symmetry
satisfy equations:
\be
{\tau_{n+1} \o \tau_{n}} \, = \, e^{\p_n} \qquad;\qquad
e^{\p_n - \p_{n-1}} \, = \,  G^n \( \bj \) \, =\, R_n          \lab{taueqs}
\ee
from which the Hirota type of Toda chain equation:
\be
\pa^2_t \ln \tau_n  = {\tau_{n+1} \, \tau_{n-1} \o \tau^2_{n}}
\lab{hirotaeq}
\ee
follows easily.
Corresponding construction for the Volterra chain objects involves both
Toda systems $(S_n,R_n)$ and $({\wti S}_n,{\wti R}_n)$ as seen from
\be
S_n - {\wti S}_{n-1} = { \pa_t B_n \o B_n} \to
B_n = e^{\int (S_n - {\wti S}_{n-1}) dt} =
e^{\p_n - {\wti \p}_{n-1}} = {\tau_{n+1} \, {\wti \tau}_{n-1} \o \tau_{n}
\, {\wti \tau}_{n}} \lab{bnppti}
\ee
while from \rf{lmiura1b} we find:
\be
A_n = {\tau_{n-1} \, {\wti \tau}_{n} \o \tau_{n} \, {\wti \tau}_{n-1}}
\lab{anppti}
\ee
This situation originates from the fact that both Toda and Volterra
chain equations were constructed in terms of the $G=g^2$ symmetry and
this symmetry ``defines" the size of the lattice spacing entering
lattice equations of motion in both \rf{aneq}, \rf{bneq} as well as
\rf{sneq}, \rf{rneq}.
It is natural to use $g$ symmetry intrinsic to the \dois to define
another connection between Toda and Volterra systems.
This introduces an idea of building the lattice system on ``finer"
spacing connected with $g$ symmetry, which would describe more naturally
the Volterra system.

\nit {\sl Dirac-Volterra equations.}
Here we use $g$ symmetry to introduce a set of (Dirac-Volterra) equations:
\be
A_{m+\h} = B_m \qquad;\qquad B_{m+\h} = A_m - { \pa_t B_m \o B_m}
\lab{dirvol}
\ee
where $m$ is now allowed to take both integer and half-integer values.
Compatibility of these two equations requires that $(A_m,B_m)$ satisfies
Volterra equations \rf{aneq}, \rf{bneq} for both half-integer and integer
values of $m$.
Hence we have two set of Volterra systems both connected to the Toda system.
The system based on integers connects to Toda via
$S_n = g^{2n} \( -\sj\, -\bsj\, + {\sj^{\,\pr} \o \sj\,}\)$,
$R_n = g^{2n} \( \bsj\, \sj\,\)$ with even powers of $g$ appearing in the
discrete Miura transformation, while the system based on half-integers
connects to  Toda via ${\wti S}_{n-1} = g^{2n-1} \( -\sj\, -\bsj\, +
{\sj^{\,\pr} \o \sj\,}\)$, ${\wti R}_{n-1} = g^{2n-1}
\( \bsj\, \sj\,\)$ with odd powers of $g$ appearing in Miura map.
These two systems constitute the total Dirac-Volterra chain consisting
of integer and half-integer sites. Each of the corresponding sub-lattices
connects independently to the ``bigger" Toda lattice.
Reproducing the Dirac-Volterra system requires therefore two Toda systems.

\nit {\sl Modified Volterra Chain.}
We will now extend our discussion of symmetry generating integrable lattice
systems to the modified Volterra (MV) equation, which has the form
\ct{W76,KM92}:
\be
\pa_t M_n = (M_n^2 -1 )  \( M_{n +1} - M_{n-1} \) \lab{mve}
\ee
MV equation connects to Volterra equation \rf{volterka} via two discrete
Miura maps:
\be
 V_n = \(1 \pm  M_n \) \( 1 \mp M_{n-1} \) \lab{mve2ve}
\ee
(for simplicity we will take from now on the upper signs in \rf{mve2ve}).
Splitting the Modified Volterra lattice as in \rf{anbnvn}
\be
L_n = M_{2n-1} \qquad;\qquad K_n = M_{2n} \lab{lnknmn}
\ee
we can rewrite \rf{mve} in components as
\br
\pa_t\, L_n = (L^2_n -1)  \( K_{n } - K_{n-1} \) \lab{lneq}\\
\pa_t\, K_n = (K^2_n-1) \( L_{n+1} - L_{n} \) \lab{kneq}
\er
while the Miura map \rf{mve2ve} takes the form:
\br
A_n \eq \, \(1 +  L_n \) \( 1 - K_{n} \) + {\pa_t L_n \o L_{n} -1}
\lab{ln2an} \\
B_n \eq \, \(1 + K_n \) \( 1 - L_{n} \) \lab{kn2bn}
\er
As a warm up exercise we first remove the constant ``one" from all the
above equations involving the MV objects.
The result is given by equation \ct{W76}:
\be
\pa_t\, M_n = M_n^2   \( M_{n +1} - M_{n-1} \) \lab{smve}
\ee
For $K_n$ and $L_n$ defined according to \rf{lnknmn} we have the Miura
relation
\br
 A_n \eq  -  L_n  K_{n-1}= -  L_{n} K_n + {\pa_t L_n \o L_n}  \lab{ln2an1} \\
 B_n \eq  -  L_{n} K_n   \lab{kn2bn1}
\er
which is of the form of \rf{j}. Associate now $\sj\,$ and $\bsj\,$
with modes $A_n$ and $B_n$ as in \rf{abn} and let $L_n =
g^{2n} (q)$ and $K_n = - g^{2n} (r)$, with $(q,r)$ being objects of
dNLS hierarchy.
Next, recall the form of the $g$ symmetry within the dNLS hierarchy.
We see that eq. \rf{smve} in components is equivalent to the shift:
\be
K_n = K_{n-1} + {\pa_t L_n \o L_n^2} \qquad;\qquad
L_{n+1} = L_{n} + {\pa_t K_n \o K_n^2}
\ee
which corresponds to the action of $g^2$ in the $(q,r)$ basis.

After this discussion we return to the complete MV equation \rf{mve}.
We still let $L_n = (\gh )^{2n} (q)$ and $K_n = - (\gh)^{2n} (r)$
with $(q,r)$ from the \treisa.
Both $L_n$ and  $K_n$ enter the MV equations \rf{lneq}, \rf{kneq}
which correspond to the lattice with spacing associated with acting with
symmetry operation $(\gh)^{2}$.
This is a reminiscent of the situation we encountered
in case of Volterra equation.
In order to associate lattice spacing to the fundamental symmetry
operation $\gh$ we define a sub-lattice including both integer and
half-integers sites and define Dirac-MV equation as
\be
L_{m+\h} = - K_m \qquad;\qquad - K_{m+\h} = L_m + { \pa_t K_m \o K^2_m-1}
\lab{dirmv}
\ee
These equations reproduce two sets of MV equations on half-integer and
integer lattices, each associated via Miura transformations with set of
Volterra variables.
\newpage
{\large {\bf 4. Remarks }}
\lskip
The main aim of this paper was to uncover connection between lattice
and continuum integrable systems.
It seems that one can view the continuum models as originating from
the much bigger structure which contains both kinematics of continuum
model as well of the lattice one.
Each site can be associated to continuum model.
In order of this structure to be consistent
one needs a symmetry which treats each site as an gauge
equivalent of the other.
Alternatively one can view existence of the discrete equations as a
consequence of the presence of this symmetry in the bigger structure.
It seems natural to think about extension of conventional description of
the integrable models in a  way which would make transparent the presence
of the extra discrete symmetry. A good point to start would be to generalize
notion of the Lax  operator in such a way that both usual and discrete
symmetries could be implemented by the gauge transformations.

It would also be interesting to see whether one finds a similar phenomena
in case of KP hierarchies build of more than two (and possibly infinitely
many) bosons.
It is tempting to suspect that this will lead to classification of all
internal KP symmetries.

Few remarks can also be made about mutual relations of
three continuum models we have considered here.
One feature which characterizes various two-boson KP hierarchies is a
number of local Poisson brackets one encounters among the multi-bracket
structures connected with the Hamiltonian structure of each model.
The model with linear form of the Lax operator, which we call
the \um possesses three local brackets ${\pbr{\cdot}{\cdot}}_i$ with
$i=1,2,3$.
The first one ($i=1$) takes the canonical form ${\pbr{X(y)}{Y(y)}}_1
= \d^{\pr} (x-y)$, with $X=J,Y=\bj$ being two conjugated objects of this
model. The remaining two structures are non-abelian.
As we now increase the non-linearity of the Lax operator passing to
the \dois we find that only the bracket structures with $i=2,3$ are local
and the canonical structure appears now for $i=2$.
The symplectic Miura map connecting \um and \dois can therefore be
understood as a Darboux map abelianizing the bracket $i=2$ of the \uma.
As we pass to the \treis the non-linearity increases and we find only one
local bracket structure for $i=3$, which by the way has the canonical form.
Hence connection between Miura maps from \um to \treis realizes the Darboux
construction for the third bracket of the \uma.
This observations are most likely partially responsible for fundamental
role played by the models we are dealing with, as one does not expect a
local bracket structures among more non-linear models, which
implies difficulties for the construction of the corresponding Hamiltonian
structures.
\lskip
{\bf Acknowledgements}
We gratefully acknowledge support within CNPq/NSF Cooperative Science Program.
One of us (HA) thanks Instituto de F\'{\i}sica Te\'{o}rica-UNESP
for kind hospitality.

\end{document}